\documentclass[epj]{svjour}
\usepackage{graphicx}
\usepackage{dcolumn}
\usepackage{bm}
\usepackage{amsmath}
\usepackage{amssymb}

\begin{document}
\title{Dynamic spin Jahn-Teller effect in small magnetic clusters}
\author{Maged Elhajal
\thanks{\emph{Present address:} Max Planck Institut, Weinberg 2, 6120 Halle, Germany}%
\and Fr\'ed\'eric Mila
}                     
\offprints{}          
\institute{Institute of Theoretical Physics, Ecole Polytechnique F\'ed\'erale de Lausanne, 1015 Lausanne, Switzerland.}
\date{Received: date / Revised version: date}
%
\abstract{
We study the effect of spin-phonon coupling in small magnetic clusters, concentrating on a S=1/2 ring of 4 spins coupled antiferromagnetically.
If the phonons are treated as classical variables, there is a critical value of the spin-phonon coupling above which a static distortion occurs. 
This is a good approximation if the zero point energy is small compared to the energy gain due to the distortion, which is true for large exchange interactions compared 
to the phonons energy ($J\gg\hbar\omega$). 
In the opposite limit, one can integrate out the phonon degrees of freedom and get an effective spin hamiltonian.
Using exact diagonalizations to include the quantum nature of both spins and phonons, we obtain the spectrum in the whole range of parameters and explicit the 
crossover between the classical and quantum regimes.
We then establish quantitatively the limits of validity of two widely used approaches (one in the quantum and one in the classical limits) and show that they are quite 
poor for small magnetic clusters.
We also show that upon reducing $\hbar\omega/J$ the first excitation of a 4-site cluster becomes a singlet, a result 
that could be relevant for Cu$_2$Te$_2$O$_5$Br$_2$.
} 
\maketitle

\section{introduction}

Although electron-phonon and spin-phonon couplings are always present in all materials, at low temperature the phonons are usually frozen and many 
properties of solids are correctly described by pure electrons and/or spins hamiltonians, as for instance the Heisenberg hamiltonian. 
There are however some examples where this coupling is primarily responsible for the low temperature physical properties. 
Superconductivity is the most famous example among electronic systems. 
In magnetic systems, the spin-Peierls instability in the antiferromagnetic spin $\frac{1}{2}$ chain \cite{Cross} and also in some frustrated two dimensionnal systems 
\cite{Federico} shows the importance that spin-phonon coupling might play. 
These instabilities with respect to a spontaneous distortion of the lattice for \emph{any} value of the spin-phonon coupling are special cases, and in standard 
systems there is a finite critical value of the spin-phonon coupling above which the system will be distorted. 
It is in particular the case for finite systems such as rings consisting of a small number of spins \cite{Spanu}.
In the case of a degenerate ground state, a distortion lifting this degeneracy is very likely to occur, which is the Jahn-Teller effect.

The phononic degrees of freedom are very often treated as classical variables. 
The atoms then have well defined positions described by a vector in real space. 
This assumption is well justified if the motion of the atoms is very slow compared to the spin dynamics, or more precisely, when the zero point energy is small 
compared to all the other energy scales. 
Going back to the spin-Peierls instability in the antiferromagnetic $S=\frac{1}{2}$ spins chain, 
this assumption will fail if the change in energy due to the distortion 
is small compared to the zero point energy of the phonons. 
In other words, if the dimerization is small, the zero point motion will restore an undistorted ground state as was shown by a quantum treatment of the phonons 
\cite{Raas,Wellein,Bursill,Augier}. 

In the following, we are interested in small magnetic rings.
This kind of systems in the presence of spin-phonon coupling was already studied with a classical treatment of the phonons \cite{Spanu}. 
In this paper, we pay special attention to the quantum nature of the phonons and show that some important departures from the classical picture may occur, depending 
on the different parameters which are the 
phonons frequency, the exchange interactions and the spin-phonon coupling.
We will concentrate on an antiferromagnetic 4-site cluster and study it in great details, treating the phonons in a fully quantum mechanical way by means of exact 
diagonalizations of the hamiltonian.
We also restrict ourselves to a ``weak'' spin-phonon coupling since this is the most relevant experimentally, but consider all possible values of the two energy 
scales involved 
in 
the problem~: the exchange interactions ($J$) and the phonons frequency ($\omega$). 
We also present a classical treatment of the phonons and show that
this approximation is valid in the limit $J\gg\hbar\omega$. 
In this limit, a spontaneous distortion is favoured since the gain in magnetic energy ($\propto J$) is much larger than the cost in elastic energy 
($\propto\hbar\omega$) and the ground state is two-fold degenerate.
Taking into account the quantum nature of the phonons  will cause some tunneling between these two classically degenerate ground states and lift the degeneracy. 
This is in complete analogy with the dynamic Jahn-Teller effect in electronic systems \cite{Moffitt}, although in our case the two-fold degeneracy is not due to a 
degeneracy of the spin system but results from the static distortion.
A semi-classical approach taking into account the tunneling between the two distorted states will be presented and the results will be compared with 
exact diagonalization to show quantitatively the limit of this method. 
In the opposite (anti-adiabatic) limit, the phonons energy is much higher than the exchange interaction and this is then the domain of validity of the pure Heisenberg 
model at low energy.
If the phonons energy is lowered, but still higher than the magnetic interations, a scheme to integrate out the phonons degrees of freedom 
\cite{Kuboki} is very often applied, which results in additionnal effective exchange interactions (renormalization of the exchange constants and appearance of 
effective longer range interactions). 
It is shown that its usefulness for small magnetic clusters is limited.
The (smooth) transition from the adiabatic ($J\gg\hbar\omega$) to the anti-adiabatic ($J\ll\hbar\omega$) regimes is shown on the basis of exact diagonalization.
Finally, a possible interpretation of the low energy spectrum of 
Cu$_2$Te$_2$O$_5$Br$_2$ is presented. 

\section{\label{sec:model}Model} %

Starting from a Heisenberg model: 
\begin{displaymath}
H_H=\sum_{i,j}J_{ij}\mathbf{S}_i.\mathbf{S}_j
\end{displaymath}
the coupling to the phononic degrees of freedom is introduced as a dependence of the coupling constants $J_{ij}(d_{ij})$ on the distances $d_{ij}$ between 
the sites $i$ and $j$. 
This is already a simplification of what a realistic coupling may be since we assume that $J_{ij}$ depends only on the distance between the atoms and not on their 
exact relative 
positions ($\mathbf{R}_i-\mathbf{R}_j$) as it does in reality. 
We are however interested in general trends of what happens in the presence of spin-phonon coupling 
and not in a fully realistic model. 
For the same reason we assume the simplest model for the motion of the atoms~: 
They form independent harmonic oscillators of frequency $\omega$. 
This is Einstein's model of dispersionless phonons.

In transition metal ions, the magnetism is due to d orbitals and the magnetic interactions are due to overlaps of these orbitals. 
In Anderson super-exchange model \cite{Anderson} (relevant for Mott insulators) the exchange interaction is $J\propto\frac{t^2}{U}$, where $t$ is the hopping between 
magnetic sites and $U$ the on-site Coulomb energy. 
The dependence of $J$ on the distance $d$ between the atoms comes through the dependence of the hopping $t$ on $d$. 
It can be shown \cite{Harrison} that for d orbitals $t\sim\frac{1}{d^5}$ which leads to an explicit expression for the spin-phonon coupling~: 
$J(d)=J_0\left(\frac{d_0}{d}\right)^\alpha$, where 
$d_0$ is the distance between the atoms at rest and $\alpha$ a number whose value is $10$ according to the previous simple arguments, but whose actual value in 
real compounds may be slightly different.
If $d$ deviates only slightly from $d_0$ an expansion of the previous expression leads to~: 
$J(d)=J_0\left(1-\alpha\frac{d-d_0}{d_0}\right)$ which will be used in the following. 

The behaviour of a cluster of spins in the presence of spin-phonon coupling depends on the number of sites and on the spin 
value \cite{Spanu}. 
However, this paper concentrates on the effects of the quantum nature of the phonons and for this purpose, 
we restrict ourselves to a ring of 4 spins where the atoms can only move along the ring. 
Let $x_i$ be the deviation of the atom $i$ with respect to its position at rest, then the hamiltonian is written as~:
\begin{eqnarray}
\label{14h59}
H&=&\frac{1}{2m}\sum_i p^{2}_{i} + \frac{k}{2}\sum_i x^{2}_{i}\\
\nonumber
&&+ J\sum_i\left[1-\frac{\alpha}{d_0}\left(x_{i+1}-x_{i}\right)\right]\mathbf{S}_{i+1}.\mathbf{S}_{i}
\end{eqnarray}
where the first two terms are the kinetic and potential energy of the phonons. 
If the phonons are treated as quantum oscillators then it is more convenient to make the usual transformations to the operators 
$a$ and $a^+$ and (\ref{14h59}) becomes~: 
\begin{eqnarray}
\label{16h49}
H&=&\hbar\omega\sum_i\left(a^{\dagger}_{i}a_{i}+\frac{1}{2}\right)\\
\nonumber
&&+J\sum_i\left[1-\beta\left(a_{i+1}+a^{\dagger}_{i+1}-a_{i}-a^{\dagger}_{i}\right)\right]\mathbf{S}_{i+1}.\mathbf{S}_{i}
\end{eqnarray}
with
\begin{equation}
\label{11h56}
\beta=\frac{\alpha}{d_0}\sqrt\frac{\hbar}{2m\omega}
\end{equation}
In the hamiltonian (\ref{16h49}), $\alpha$, $d_0$ and $m$ are microscopic parameters and in the 
following they will be fixed to some typical value. 
Although there are only two energy scales ($J$ and $\hbar\omega$), it is \emph{not} possible to reduce the parameters in 
(\ref{16h49}) to only one relevant parameter (for example 
$\frac{\hbar\omega}{J}$) because $\beta\propto\frac{1}{\sqrt{\omega}}$ as is seen from (\ref{11h56}). 
However, to avoid dealing with two independent parameters,
in the following $\omega$ is fixed to a typical value of $\frac{\hbar\omega}{k_B}=100 K$ which is a realistic value for phonons in real compounds, and $J$ 
will be varied.
For this value of $\omega$, taking typical values $\alpha\sim10$, $d_0\sim3.5$ \AA \ and $m\sim$60 atomic masses, we find 
$\beta\sim0.2$. 
Except when specified, $\beta$ will be fixed to this typical value in the following. 
For other values of $\beta$ the physical pictures presented remain valid although the results may depend quantitatively (but not qualitatively) on 
$\beta$.
An example of this dependence is investigated in more detail in section \ref{physical_applications} where $\beta$ is varied.

Strictly speaking, one cannot express all the results as a function of the
ratio $\frac{J}{\hbar\omega}$, because $\beta$ depends also on $\omega$ and the calculations are actually done for one fixed value of $\omega$.
However, keeping in mind that we have fixed $\frac{\hbar\omega}{k_B}=100 K$, we will express
all the results as a function of $\frac{J}{\hbar\omega}$ because the relative
values of these two energy scales \emph{is} the relevant parameter which determines the characteristic
behaviour of the system.

\section{\label{17h33}Exact diagonalizations.} %

In order to keep the quantum nature of the phonons we 
will diagonalize exactly the hamiltonian (\ref{16h49}). 
We first give some technical details about exact diagonalizations, which can be 
skipped by the uninterested reader.

\subsection{Technical details}

\begin{figure}
\begin{center}
\resizebox{0.75\columnwidth}{!}{%
\includegraphics[width=9cm]{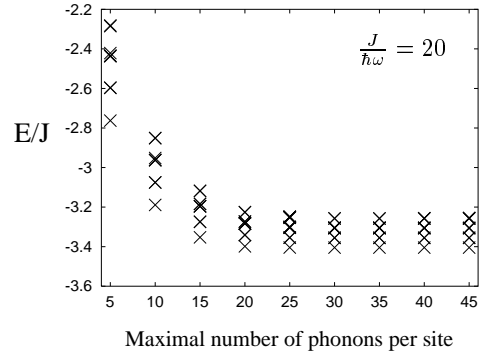}
}
\caption{\label{conv_1}Evolution of the low energy spectrum with the maximum
number of phonons allowed on each site. Symmetries are used, and
only the lowest energy levels corresponding to the irreducible representation ($S_z=0$, $k=0$) are shown.}
\end{center}
\end{figure}

\begin{figure}
\begin{center}
\resizebox{0.75\columnwidth}{!}{%
\includegraphics[width=9cm]{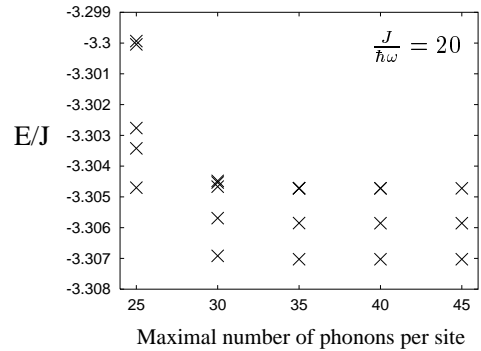}
}
\caption{\label{conv_2}Detailed view of the third group of levels from figure \ref{conv_1}. 
Although the rough convergence seems to be achieved for $30$ phonons per site (and indeed the ground state energy is already converged to $\sim 10^{-6}$), more 
phonons are needed to get the fine structure of the spectrum (notice the fake degeneracy lifting for $30$ phonons per site).}
\end{center}
\end{figure}

The dimension of the Hilbert space of a quantum harmonic oscillator is infinite and it should be truncated in order to 
be diagonalized numerically. 
This is done by introducing a cut-off in the maximal number of phonons on each site. 
For each set of parameters, we estimated the error introduced by this cut-off by inspection of the 
evolution of a subset of energy levels as the cut-off is varied (see figures
\ref{conv_1} and \ref{conv_2}). 
For all the results presented in the following sections, the error bars would be completely negligible and for this reason they were omitted.
The phonon wave function was described using the basis in which the number of phonons is diagonal. 
This is a good choice in the limit of $\hbar\omega\gg J$ since the number of phonons is then 
``almost'' a good quantum number (see section \ref{17h28}). 
The convergence, with respect to the number of phonons states taken into account, is then very fast because the eigenstates will be correctly approximated using a 
small number of vectors from the 
choosen basis. 
Indeed, for the largest values of $\hbar\omega$ considered in this work, it was found that taking into account a 
maximum number of 5 phonons per site is 
already an excellent approximation.
As $\hbar\omega$ is decreased, more and more excited levels of the harmonic oscillators are involved. 
In principle, to build the (classical) static distortion found in section \ref{17h29}, 
all the energy levels of the harmonic oscillators are needed. 
For the largest values of $J$ considered, up to 40 phonons 
per site were taken into account in order to get a good convergence of the results. 
This represents a huge Hilbert space and it was necessary to take into 
account both geometrical symmetries of the cluster and the fact that $S_z$ is a good quantum 
number in order to be able to diagonalize numerically the hamiltonian, using Lanczos's algorithm. 
The diagonalization was performed separately in each of the $S_z=0,1 \text{ and } 2$ subspaces ($S_z=-1 \text{ and } -2$ being respectively degenerate with 
$S_z=1 \text{ and } 2$).
The geometrical symmetries we took into account are the 4 rotations (including identity) around the axis passing through the middle of the ring and perpendicular to 
it. 
These rotations perform circular permutations of the 4 sites and if the ring is seen as a linear chain with periodic boundary conditions then they are equivalent to 4 
translations along the chain.
There are 4 irreducible representations for this group of symmetries which we label with $k=0,1,2 \text{ and } 3$.
The diagonalizations are performed separately in each of the subspaces associated with these 4 irreducible representations.

\subsection{Results}

\begin{figure}
\begin{center}
\resizebox{0.75\columnwidth}{!}{%
\includegraphics[width=9cm]{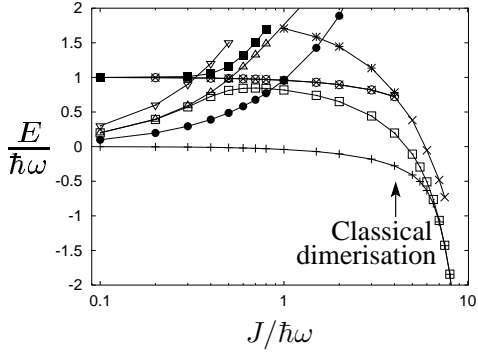}
}
\caption{\label{fig6}Transition from the anti-adiabatic ($\hbar\omega\gg J$) to the adiabatic ($\hbar\omega\ll J$) regimes. 
For each value of $\frac{J}{\hbar\omega}$ only the lowest states are represented. 
In the anti-adiabatic limit, the low-energy spectrum is made of the four levels of the pure Heisenberg hamiltonian. 
As $\frac{J}{\hbar\omega}$ increases, initially higher energy levels involving phonon degrees of freedom come into play. 
For large values of $\frac{J}{\hbar\omega}$, the system enters the adiabatic regime where the phonons behave classically. 
The ground state is then two-fold degenerate, since one singlet state (empty squares) falls on the ground state after an anti-crossing with a level of higher energy 
(full squares). 
For comparison, the critical value of $\frac{J}{\hbar\omega}$ above which a static dimerization is predicted in a classical treatment of the phonons is shown. 
}
\end{center}
\end{figure}

The evolution of the spectrum as a function of $\frac{J}{\hbar\omega}$ obtained from exact diagonalizations is shown in figure \ref{fig6}.
Several aspects of this spectrum will be studied in more details in the next sections and only a few remarks will be made here. 

First of all, for small $\frac{J}{\hbar\omega}$ where the phonons are
essentially frozen at low energy, one can recognize the four levels of the
pure Heisenberg model (without coupling to the phonons). 
In the opposit limit (large $\frac{J}{\hbar\omega}$) it is seen that the ground state is two-fold degenerate, which is the quantum equivalent of the static
distortion predicted when the phonons are treated classically (see figure \ref{fig5}). 
However, the value of $\frac{J}{\hbar\omega}$ at which the distortion appears
in the classical approximation (arrow on figure \ref{fig6})
is quite different from that at which the degeneracy of the ground state appears (see section \ref{17h29} for
more details). 
The continuous evolution of the spectrum from one limit to the other is one of the 
original results of our work. 
Several methods exist to solve the problem starting from one limit or the
other. 
They will be presented in the next sections, which will help understand the spectrum of figure \ref{fig6} and the domains of validity of these approximate
methods will be made explicit by comparison with the results of exact diagonalizations.

\section{\label{17h29}Classical (adiabatic) and semi-classical phonons} %

Small rings of spins with adiabatic phonons were studied and several interesting features were found by varying the 
number of sites and the spin \cite{Spanu}. 
The calculation for the 4-site $S=\frac{1}{2}$ ring is repeated here for 
the sake of comparison between the classical (adiabatic), semi-classical and fully quantum approaches in the next paragraphs. 

\subsection{Classical phonons}

\begin{figure}
\begin{center}
\resizebox{0.75\columnwidth}{!}{%
\includegraphics[height=2cm]{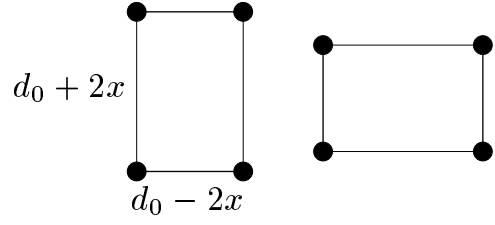}
}
\caption{\label{fig5}Two degenerate ground states obtained with classical (adiabatic) phonons if the spin-phonon coupling is above the critical value (equation 
\ref{11h55}).}
\end{center}
\end{figure}

The ground state of the hamiltonian (\ref{14h59}) is achieved when the atoms are at rest ($\mathbf{p}_i=\mathbf{0}$) and at the minimum of the total energy~:
$$
E=\frac{k}{2}\sum_i\mathbf{x}^2_i+E_M(\mathbf{x}_i)
$$
where $E_M$ is the energy of the spin system ground state for fixed values of the positions $\mathbf{x}_i$. 
$E_M$ is calculated quantum-mechanically.
Minimizing this expression with respect to the $\mathbf{x}_i$ yields the energy of the ground state~:
$$
E=-J-J\sqrt{1+3\left(\frac{2\alpha}{d_0}\right)^2x^2}+2kx^2
$$
with $x$ the displacement as represented on figure \ref{fig5}.
This expression has a minimum at finite $x$ only for large enough spin-phonon coupling constants.
The critical value is~:
\begin{equation}
\label{11h55}
\alpha_c=d_0\sqrt{\frac{k}{3J}}
\end{equation}
For $\alpha<\alpha_c$ the atoms will remain undisplaced, whereas for $\alpha>\alpha_c$ a static distortion, as represented on figure \ref{fig5}, will occur.
The amplitude of the distortion is then~:
\begin{equation}
\label{11h57}
x=\frac{1}{2}\sqrt{3\left(\frac{\alpha J}{kd_0}\right)^2-\frac{1}{3}\left(\frac{d_0}{\alpha}\right)^2}
\end{equation}

\subsection{Semi-classical approach}

The preceeding classical method neglects the quantum fluctuations of the phonons and it can be improved by reincorporating them on top of the classical solution. 
In order to do this, we first solve the quantum problem in a mean-field approximation inspired by the classical dimerization of figure \ref{fig5}.
This is done by introducing ``by hand'' a static distortion~:
\begin{equation}
\label{15h00}
\delta=\langle a_1+a^+_1\rangle=\langle a_3+a^+_3\rangle=-\langle a_2+a^+_2\rangle=-\langle a_4+a^+_4\rangle
\end{equation}
which plays the role of the classical variable $x$ in the classical treatment since for a harmonic oscillator 
$\langle x\rangle=\sqrt{\frac{\hbar}{2m\omega}}\langle a+a^+\rangle$.
The two distorted states of figure \ref{fig5} correspond to $\delta>0$ and $\delta<0$.
Using (\ref{15h00}), we replace in the hamiltonian (\ref{16h49}) the $a_i$ and $a^+_i$ by their expression in term of $\delta$.
Doing so, one obtains a hamiltonian with only spin variables which can be diagonalized exactly, 
and the mean values of $\langle\mathbf{S}_i.\mathbf{S}_{i+1}\rangle$ in the ground state is calculated as a function of 
$\delta$.
Replacing the $\mathbf{S}_i.\mathbf{S}_{i+1}$ by their mean value in the hamiltonian (\ref{16h49}), one obtains then a hamiltonian which depends on $\delta$ 
with only phonon variables.
This is a displaced harmonic oscillator hamiltonian which is easily diagonalized and finally, 
self-consistency 
is obtained by requiring that this phononic hamiltonian leads to the same distortion $\delta$ as the one first introduced in the spin hamiltonian. 

Not surprisingly, the results are similar to the adiabatic phonons results. The distortion $\delta$ and the critical value $\beta_c$~: 
\begin{equation}
\label{12h34}
\delta=\pm\sqrt{3\left(\frac{\beta J}{\hbar\omega}\right)^2-\frac{1}{12\beta^2}} \quad \quad \beta_c=\sqrt{\frac{\hbar\omega}{6J}}
\end{equation}
are the same as (\ref{11h57}) and (\ref{11h55}) given the relationship between $\alpha$ and $\beta$ (see equation (\ref{11h56})) and between $\delta$ and $x$ 
(see discussion after equation (6)).
However, unlike in the pure classical approach, we have now an explicit wave function of the ground state which will be a direct product of a phonon part 
and a spin wave function. 
The phonon part is the wave function of a displaced harmonic oscillator, with the displacement given by equation (\ref{12h34}). 
The spin wave function is~:
$$
\begin{array}{l}
\frac{1}{\sqrt{6}}\sqrt{1+Y}\left(|\uparrow\downarrow\uparrow\downarrow\rangle+|\downarrow\uparrow\downarrow\uparrow\rangle\right)+\\
+\left(-\frac{1}{2\sqrt{6}}\sqrt{1+Y}\pm\frac{1}{2\sqrt{2}}\sqrt{1-Y}\right)\left(|\uparrow\uparrow\downarrow\downarrow\rangle+|\downarrow\downarrow\uparrow\uparrow\rangle\right)+\\
+\left(-\frac{1}{2\sqrt{6}}\sqrt{1+Y}\mp\frac{1}{2\sqrt{2}}\sqrt{1-Y}\right)\left(|\uparrow\downarrow\downarrow\uparrow\rangle+|\downarrow\uparrow\uparrow\downarrow\rangle\right)
\end{array}
$$
where $Y=\frac{\hbar\omega}{6J\beta^2}$ and the two possibilities ($\pm$) correspond to positive and negative $\delta$ (the two solutions of figure \ref{fig5}).

This mean-field solution has neglected the quantum fluctuations, and one can improve it by re-introducing them, on top of 
the two-fold degenerate distorted mean-field ground states. 
Technically, this is achieved by restricting the Hilbert space to a two-dimensionnal space generated by the two solutions obtained above. 
Rewriting the full hamiltonian (\ref{16h49}) in this subspace, and diagonalizing it, one obtains two energy levels~:
\begin{eqnarray}
\label{12h00}
E_0&=&\frac{a+b}{1+\langle 1|2\rangle}\\
\label{11h11}
E_1&=&\frac{a-b}{1-\langle 1|2\rangle}
\end{eqnarray}
where
\begin{eqnarray}
a&=&2\hbar\omega-\frac{J}{2}\left(2+Y+\frac{1}{Y}\right)\\
\label{10h23}
b=\langle 1|H|2\rangle\hspace{-2mm}&=&\hspace{-2mm}\left[-\frac{J}{2}\left(3+2Y-Y^2\right)+2Y\hbar\omega\right]e^{-2\delta^2}\label{12h12}\\
\label{10h21}
\langle1|2\rangle&=&Ye^{-2\delta^2}
\end{eqnarray}
$\langle1|2\rangle$ being the overlap between the two mean-field solutions (the exponential comes from the overlap of the phonon wave functions 
and $Y$ from the overlap of the spin wave functions). 
It is useful for the following discussion to note that $b$ is the matrix element of the hamiltonian connecting the two states and the splitting between them 
is~:
 \begin{equation}
E_1-E_0=\frac{2Je^{-2\delta^2}\left(1-Y^2\right)}{1-Y^2e^{-4\delta^2}}
\end{equation}

\subsection{Discussion}

It is easily seen on the expression (\ref{12h34}) of the distortion $\delta$ that for large values of the ratio $\frac{J}{\hbar\omega}$ the distortion will be 
important (which is intuitive since the gain in magnetic energy is large while the elastic cost is small). 
As $\frac{J}{\hbar\omega}$ and $\delta$ increase, the overlap (\ref{10h21}) and the matrix element (\ref{10h23}) between the two states will vanish exponentially. 
The two mean-field solutions (identical to the classical result) will thus become assymptotically eigenstates of the hamiltonian as $\frac{J}{\hbar\omega}\to\infty$. 
This is how the classical behaviour of the phonons is recovered starting from the quantum hamiltonian. 
As $\frac{J}{\hbar\omega}$ is lowered, the matrix element between the two states will become more and more important, giving rise to a tunnelling probability between 
the two classical states and a splitting will appear. 

\begin{figure}
\begin{center}
\resizebox{0.75\columnwidth}{!}{%
\includegraphics[width=9cm]{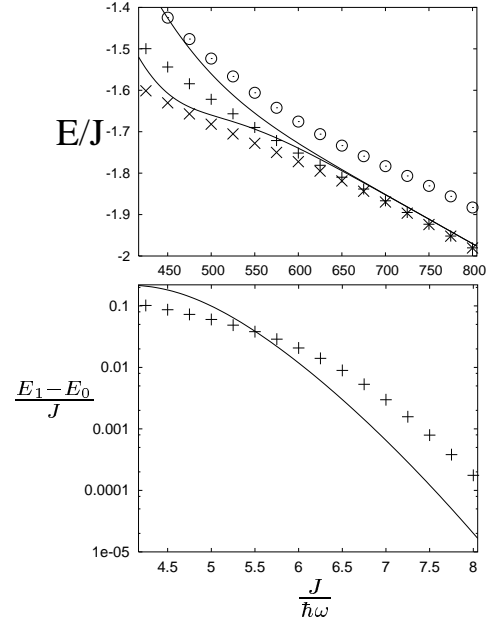}
}
\caption{\label{fig4}Semi-classical approach from the adiabatic limit. 
(top) Comparison of the energies of the two lowest levels (equations (\ref{12h00}) and (\ref{11h11})) as obtained from the semi-classical approach 
(solid line) and exact diagonalization (points). 
For larger values of $\frac{J}{\hbar\omega}$ the energies obtained with both methods will become identical (not shown here for clarity). 
The third energy level from exact diagonalization was included. 
(bottom) Comparison of the splitting between the two lowest energy levels from the semi-classical approach and exact diagonalization. 
Although the agreement is not excellent, the same exponential behaviour appears
for large $\frac{J}{\hbar\omega}$ in both approaches. 
}
\end{center}
\end{figure}

The two energy levels obtained, as well as the splitting between the two states, are represented on figure \ref{fig4} and compared with exact diagonalizations. 
As seen from the figure, the approach is qualitatively correct since for large $\frac{J}{\hbar\omega}$, the exact spectrum has two energy levels well separated from 
the higher levels and which show the same behaviour as predicted by the semi-classical approach. 
The semi-classical approach however fails to predict quantitatively the value of the splitting although it has the correct exponential behaviour 
at large $\frac{J}{\hbar\omega}$.

\begin{figure}
\begin{center}
\resizebox{0.75\columnwidth}{!}{%
\includegraphics[height=7cm]{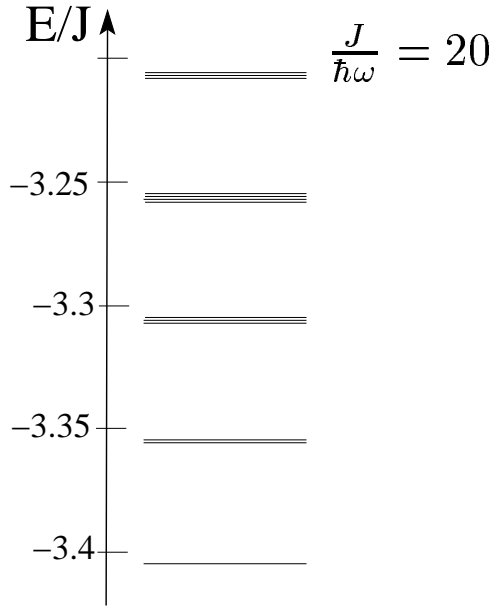}
}
\caption{\label{fig1}Low energy spectrum obtained with exact diagonalization in the adiabatic regime ($J\gg\hbar\omega$).
All states are singlet states, triplet states being much higher in energy.
The ground state is two fold degenerate, which is the quantum equivalent of the two classically dimerized states.
The rough structure of the excited states correspond to harmonic vibrations around 
the dimerized states.
The splitting within one group of close lying states can be seen as a perturbation of these oscillations by the magnetic degrees of freedom.}
\end{center}
\end{figure}

Finally, in the regime where $\frac{J}{\hbar\omega}$ is very large, a very
simple physical picture arises. 
The system is expected to be classically dimerized and the low energy degrees
of freedom are the vibrations around the distorted states. 
They are thus displaced harmonic oscillators with the same frequency as the
phonons. 
This scenario is confirmed by exact diagonalizations (see figure \ref{fig1})

\section{\label{17h28}Expansion from the anti-adiabatic limit} %

In the abscence of spin-phonon coupling the eigenstates of the hamiltonian are direct products of a magnetic and phononic 
parts.
Furthermore, if $\hbar\omega\gg J$, the energy levels will be grouped around the energy levels of the harmonic oscillators (which are the large energy scale) and 
within each group there will be a fine structure given by the magnetic part. 
The lowest group of states is nothing else but the spectrum of the pure Heisenberg model, since these states have no phonon. 
Including spin-phonon coupling, one can consider this term as a perturbation compared to the pure Heisenberg and the pure 
phononic terms. 
This is because $\beta$ is a rather small number and because the lowest (unperturbed) states 
have zero phonons and thus the phononic operators $a_i$ and $a_i^+$ will give contributions of the order of 
$\frac{1}{\hbar\omega}$ to second order in perturbation. 
Restricting the attention to these zero phonon states, one can take into account the spin-phonon coupling in 
perturbation and interpret the result as an effective spin hamiltonian, thus integrating out the phonons degrees of 
freedom \cite{Kuboki}.
To second order in perturbation theory, this will introduce effective second neighbours interactions and renormalize the 
first neighbours interactions, so that the 
total hamiltonian has the usual Heisenberg part and in addition~:
\begin{equation}
\label{18h20}
\sum_i\frac{J^2\beta^2}{\hbar\omega}\left(\mathbf{S}_{i+1}.\mathbf{S}_i+\frac{3}{2}\mathbf{S}_{i+2}.\mathbf{S}_i-\frac{3}{8}\right)
\end{equation}
Two distinct assumptions must be fulfilled in order for this effective hamiltonian to be valid, although they are both 
related to the fact that $\hbar\omega\gg J$.
The first one is the usual validity of perturbation theory, and one could improve the above 
scheme by going to higher orders in perturbation. 
The second assumption is that the phononic degrees of freedom (which are integrated out) 
remain at higher energies than the spectrum obtained with the above effective hamiltonian. 
In other words, the states which have one or more phonons should remain well 
separated from the lowest energy levels for which we derived the effective hamiltonian. 

\begin{figure}
\begin{center}
\resizebox{0.75\columnwidth}{!}{%
\includegraphics[width=7cm]{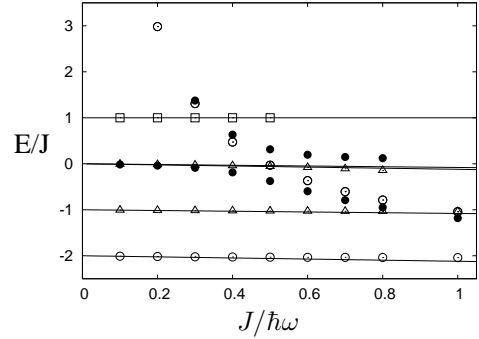}
}
\caption{\label{fig2}Low energy spectrum as a function of $\frac{J}{\hbar\omega}$.
The lines are the spectrum obtained with a pure Heisenberg model renormalized by high energy phonons to second order in perturbation 
(see text and equation (\ref{18h20})).
The points are obtained from exact diagonalization of the hamiltonian (\ref{16h49}).
For clarity only the lowest energy levels were represented at each value of $J/\hbar\omega$, and an energy of $2\hbar\omega$ 
(zero point energy of the 4 phonon modes) was substracted. 
S=0 states are represented by circles, S=1 states by triangles and S=2 states by squares. 
The anti-crossing between two S=0 states (full circles) is emphasized.
The lowest of these two states will be the first excitation below the first triplet state for $\frac{J}{\hbar\omega}\gtrsim 0.9$.
In the adiabatic limit ($J\gg\hbar\omega$) it will fall onto the ground state (see figure \ref{fig6}).
}
\end{center}
\end{figure}

Exact diagonalization results are compared with the energy levels of the pure Heisenberg hamiltonian, 
renormalized by (\ref{18h20}) (see figure \ref{fig2}). 
For very small $\frac{J}{\hbar\omega}$, the low energy spectrum is essentially that of a pure Heisenberg hamiltonian (4 initial 
levels for small $\frac{J}{\hbar\omega}$), 
and the perturbative approach reproduces well the small departures ($\sim 1\%$) as $\frac{J}{\hbar\omega}$ increases. 
For $\frac{J}{\hbar\omega}\sim 0.4$, states from higher energy involving phonons will fall on the lowest levels, causing great departures from the (renormalized) 
Heisenberg model. 
The perturbative approach fails completely to reproduce such a change in the spectrum since it integrates out the phonons.
This commonly used method (for other systems) appears thus useless for small magnetic systems since, in its range of validity it only predicts departures of 
$\sim 1\%$ in the renormalized exchange constants, which is orders of magnitude smaller than the perturbation brought by higher energy phonons as seen in exact 
diagonalizations.

\section{\label{physical_applications}Physical applications} %

Appart from testing the validity of various approximate methods, two aspects should be underlined from the results of the previous sections, which are important for small magnetic clusters, such as 
molecular magnets.
First, the spin-phonon coupling can tremendously change the low energy spectrum of small magnetic clusters. 
This was already pointed out by a previous work where the phonons were treated classically \cite{Spanu}.
Second, the quantum fluctuations associated to the phononic degrees of freedom must be taken into account to describe correctly the low energy spectrum, except for the classical limit ($J\gg\hbar\omega$), 
which is unlikely to occur in molecular magnets, given the smallness of the exchange interactions in these systems, which are typically much smaller than the phonon frequencies. 
It was also shown that the spin-phonon degrees of freedom will qualitatively change the low energy spectrum in small magnetic clusters, even in the abscence of a static distortion, a phenomenon which is 
similar to the dynamic Jahn-Teller effect for the electronic degrees of freedom.

For instance, it was shown how a spin-phonon coupling can be responsible for a singlet-singlet first excitation in a 4-site cluster (whereas the lowest transition is a singlet-triplet one in the absence 
of spin-phonon coupling). 
The value of the spin-phonon coupling was so far fixed to $\beta=0.2$ and to obtain a singlet first excitation requires $\frac{J}{\hbar\omega}\gtrsim 0.9$.
Higher values of the spin-phonon coupling would however require lower values of $\frac{J}{\hbar\omega}$ which is more likely to occur in real materials.
In order to make this argument more quantitative, figure \ref{fig8} shows the dependence on $\frac{J}{\hbar\omega}$ of $\beta_{SS}$, the critical value of $\beta$ 
needed for the first excited singlet to be lower in 
energy than the first triplet. 
As can be seen from this figure, even if the phonon frequencies are up to two times the exchange interactions, reasonable values of $\beta$ are expected to be enough 
to 
change 
qualitatively the spectrum and to bring a singlet excitation below the first triplet.

The magnetic properties of the compound Cu$_2$Te$_2$O$_5$Br$_2$ are not fully understood. 
Although magnetic measurements \cite{Lemmens} and ab-initio calculations \cite{Vladimir} point to a system of weakly coupled plaquettes of 4 spin-$\frac{1}{2}$ copper ions, a magnetic order was 
recently observed \cite{Zaharko}. 
Partial explanations were given \cite{Gros,Valeri} but an overall understanding of all the features is still lacking, 
the main puzzle being the existence of both magnetic order and a low lying singlet excitation.
A coupling between the plaquettes is needed for the appearance of magnetic order and the observed phase transition. 
However, some features, such as the existence of a low lying singlet excitation can be understood on the level of one isolated plaquette. 

\begin{figure}
\begin{center}
\resizebox{0.75\columnwidth}{!}{%
\includegraphics[height=2cm]{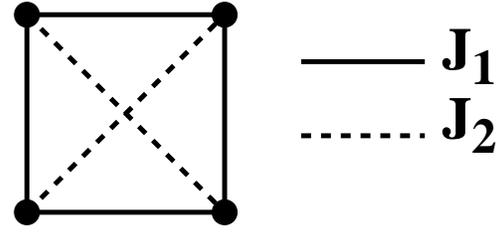}
}
\caption{\label{fig3}Elementary plaquette of four spin-$\frac{1}{2}$ Cu$^{2+}$ ions in Cu$_2$Te$_2$O$_5$Br$_2$ with first ($J_1$) and second $J_2$ neighbours 
couplings. 
ab-initio calculations find negligible $J_2$, whereas Raman spectroscopy shows a first singlet excitation which implies significant $J_2$ in a pure Heisenberg model. 
It is proposed that indeed $J_2$ is negligible and the first singlet excitation is due to spin-phonon coupling
}
\end{center}
\end{figure}

\begin{figure}
\begin{center}
\resizebox{0.75\columnwidth}{!}{%
\includegraphics[width=7cm]{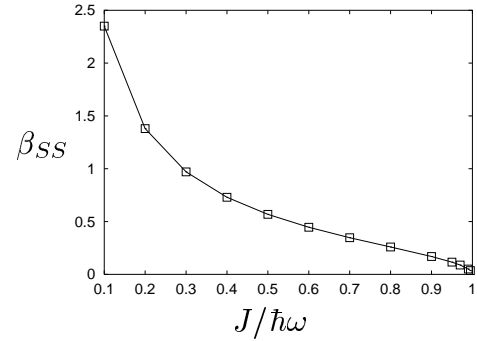}
}
\caption{\label{fig8}Critical value of the spin-phonon coupling $\beta$ needed in order to bring the first singlet excited state below the first triplet.}
\end{center}
\end{figure}

Experimentally, Raman spectroscopy \cite{Lemmens} showed that the ground state as well as the first excitation are singlet states. 
Due to the symmetry of the cristal, only two coupling constants can appear in one plaquette, as represented on figure \ref{fig3}. 
If $J_2$ is of the order of $J_1$ then the ground state, as well as the first excited state will be singlet (S=0) states. 
However, ab-initio calculations have shown that $J_2$ is in fact negligible \cite{Vladimir}, in which case the first excitation is a 
triplet state and the first singlet excitation 
is twice higher in energy, in contradiction with experimental results. 
We propose that spin-phonon coupling may be responsible for a first singlet excitation, in the abscence of second nearest neighbours coupling ($J_2$).
Figure \ref{fig6} indeed shows that for $J\gtrsim\hbar\omega$, the first excited states are two almost 
degenerate singlet states, below the first triplet state. 
The singlet excitation is actually not a pure magnetic excitation, as would be obtained by a next-nearest neighbour coupling $J_2$, since 
it involves also phononic degrees of freedom. 
Although the present model is over-simplified for a quantitative description of the compound Cu$_2$Te$_2$O$_5$Br$_2$, it shows that such a mechanism is possible.

\section{Conclusions} %

We have studied the effect of spin-phonon coupling in a 4-site $S=\frac{1}{2}$ antiferromagnetic ring taking into account exactly the quantum nature of the phonons 
by means of exact diagonalizations. 
Perturbative and variationnal approaches in the adiabatic and anti-adiabatic limits were presented and compared to exact diagonalizations and a quantitative estimate 
of their domain of validity and accuracy was presented. 
The evolution of the spectrum from the classical regime to the anti-adiabatic one was shown by means of exact diagonalizations. 
From these results, it is expected that spin-phonon coupling might play an important role in small magnetic clusters, as for instance in molecular magnets. 
Furthermore, it is very likely that taking into account the quantum fluctuations of the phonons is necessary to achieve a correct description of these systems. 
This is examplified in the magnetic analog of the dynamic Jahn-Teller effect which was underlined.
An interpretation of the low energy spectrum of Cu$_2$Te$_2$O$_5$Br$_2$ was proposed. 

We have concentrated on a 4-site ring, but we believe the results are valid for other clusters, at least qualitatively.
This is supported by the similarity of even clusters if the phonons are treated classically \cite{Spanu}.
There might however exist a discrepancy between clusters with even and odd numbers of sites.
Indeed, even-rings have a 2 fold degenerate distorted ground state, whereas this pattern is frustrated in the case of odd clusters, leading to a much higher (classical) degeneracy. 
The quantum fluctuations might thus play an even more important role in odd clusters.
Another significant parameter could be the value of the spin, since this dependence was observed in clusters with a classical treatment of the phonons \cite{Spanu}.
These problems are left for future investigation.

\subsection*{Aknowledgments} %

We thank V. Kotov and G.S. Uhrig for useful discussions and MaNEP and the Swiss National Fund for support.

\end{document}